\documentclass[conference]{IEEEtran}
\IEEEoverridecommandlockouts
\usepackage{cite}
\usepackage{amsmath,amssymb,amsfonts}
\usepackage{algorithmic}
\usepackage{graphicx}
\usepackage{textcomp}
\usepackage{xcolor}
\usepackage{gensymb}

\def\BibTeX{{\rm B\kern-.05em{\sc i\kern-.025em b}\kern-.08em
    T\kern-.1667em\lower.7ex\hbox{E}\kern-.125emX}}
\begin{document}

\title{Deep Open Space Segmentation using\\ Automotive Radar}
%\thanks{Identify applicable funding agency here. If none, delete this.}

\author{\IEEEauthorblockN{1\textsuperscript{st} Farzan Erlik Nowruzi}
\IEEEauthorblockA{\textit{University of Ottawa} \\
Ottawa, Canada \\
fnowr010@uottawa.ca}
\and
\IEEEauthorblockN{2\textsuperscript{nd} Dhanvin Kolhatkar}
\IEEEauthorblockA{\textit{Sensor Cortek Inc.} \\
Ottawa, Canada \\
dhanvin@sensorcortek.ai}
\and
\IEEEauthorblockN{3\textsuperscript{rd} Prince Kapoor}
\IEEEauthorblockA{\textit{Sensor Cortek Inc.} \\
Ottawa, Canada \\
prince@sensorcortek.ai}
\and
\IEEEauthorblockN{4\textsuperscript{th} Fahed Al Hassanat}
\IEEEauthorblockA{\textit{Sensor Cortek Inc.} \\
Ottawa, Canada \\
fahed@sensorcortek.ai}
\and
\IEEEauthorblockN{5\textsuperscript{th} Elnaz Jahani Heravi}
\IEEEauthorblockA{\textit{Sensor Cortek Inc.} \\
Ottawa, Canada \\
elena@sensorcortek.ai}
\and
\IEEEauthorblockN{6\textsuperscript{th} Robert Laganiere}
\IEEEauthorblockA{\textit{University of Ottawa} \\
Ottawa, Canada \\
laganier@eecs.uottawa.ca}
\and
\IEEEauthorblockN{7\textsuperscript{th} Julien Rebut}
\IEEEauthorblockA{\textit{Valeo} \\
%\textit{University of Ottawa}\\
Paris, France \\
julien.rebut@valeo.com}
\and
\IEEEauthorblockN{8\textsuperscript{th} Waqas Malik}
\IEEEauthorblockA{\textit{Valeo} \\
%\textit{University of Ottawa}\\
Bietigheim-Bissingen, Germany \\
waqas.malik@valeo.com}
}

\maketitle

\begin{abstract}
%Camera and lidar processing have been revolutionized with the rapid development of deep learning model architectures. 
%Automotive radar is one of the crucial elements of automated driver assistance and autonomous driving systems. Radar still relies on traditional signal processing techniques unlike Camera and Lidar based methods. We believe this is the missing link to achieve the most robust perception system. Identifying drivable space and occupied space is the first step in any autonomous decision making task. Occupancy grid map representation of the environment is often used for this purpose. 
In this work, we propose the use of radar with advanced deep segmentation models to identify open space in parking scenarios. A publically available dataset of radar observations called SCORP was collected. Deep models are evaluated with various radar input representations. Our proposed approach achieves low memory usage and real-time processing speeds, and is thus very well suited for embedded deployment.
\end{abstract}


\section{Introduction}
\label{introduction}

Levels of autonomy in driving systems are categorized between level 1 driver assistance systems to fully autonomous level 5 systems. 
%Autonomous driving can be approached from various sensor modalities such as camera, radar, sonar, or lidar.
%	Each sensor provides a piece of valuable information. It is shown that the fusion of multiple sensors is required to cover all of the environmental variations and achieve fully autonomous driving. Despite this observation, each individual sensor needs to be pushed to the edge of its capabilities to reduce the complexity of the fusion systems.
%The first systems were mostly used in the premium car segment for comfort applications, such as Adaptive Cruise Control (ACC). With the continuous improvement of radar technology, recent radar sensors are also used in safety applications.
%, such as Autonomous Emergency Braking (AEB). 
Radar is a low-powered sensor which has been used in the automotive industry for a few decades, offering a less expensive depth estimation solution than lidar. They are a crucial component of autonomous vehicles due to their capability to observe objects and their instantaneous velocities, and their robustness in harsh weather conditions.
%Furthermore, radars are not affected by poor lightning and fog, unlike camera and lidar respectively. The new versions of radar will significantly increase the resolution of the observations, therefore enabling an even wider variety of applications.

There are numerous deep learning model architectures which show state-of-the-art performance on various computer vision applications~\cite{2018_deeplabv3+, FCN}, but they are generally used with input sensors such as camera and lidar.
%However, these model architectures rely on camera and lidar as their input sensors. 
%Radar is an alternative depth sensor that presents a less expensive solution than lidar. 
Despite its capabilities, radar still depends on traditional signal processing techniques, and is seldom used with modern deep learning methods. This could be attributed to the unintuitive nature of its information representation, and to the lack of publicly available datasets.
%There are many possible signal configurations, depending on parameters such as the number of transmitters, receivers, virtual channels, samples, and chirps, the shape of these signals, mode of transmission, \textit{etc.}, that need to be adjusted to capture the radar signal. 
Once radar echoes are collected, they require various conversions from time domain to frequency domain to be understandable by human experts. Furthermore, annotating these signals is a challenging task that requires careful consideration.

%In this paper, we introduce a deep-learning based open space segmentation model which operates on various levels of radar signal representation in parking lot scenarios.

%various sub-systems required to capture, process, and use the Radar data. These sub-systems include the signal configuration, intrinsic, extrinsic, and multi-sensor calibration, and various aspects of Radar data processing pipeline. We further demonstrate how the neural models can be used to enhance the performance of a Radar based systems.

%In this paper, we collect a dataset of raw radar observations and propose a deep learning model architecture that trains on processed radar signals in both the frequency and the spatial domain. The objective of the model is to output a free space segmentation map of a parking environment on a per-frame basis. 
The contributions presented in this paper are as follows: 
\begin{enumerate}
	\item The first publically available comprehensive dataset including raw radar inputs along with ground-truth open space annotations.
	\item A deep segmentation approach that consumes radar signals to estimate open space in a parking lot. Our proposal provides comparable performance to much larger models while being faster and smaller.
	\item A comprehensive study of various radar modalities for the purpose of open area segmentation.
	%Evaluation of various domains of radar input and label representation, studying the necessity of .
	%\item Evaluation of various models and loss functions for this task.
	\item Evaluation of multiple deep learning approaches on the collected dataset.
\end{enumerate}

%	The literature review of deep segmentation model architectures along with radar applications are discussed in Section~\ref{litrev}. Section~\ref{dataset} explains the data capture setup and radar processing pipeline. The dataset, parameters, and the annotation process is also outlined in this section. The compared model architectures are  described in Section~\ref{model}. In Section~\ref{exps}, various experiments including the effect of different input data representations, comparative study of various models, and the effect of various loss functions on model performance are evaluated.

\section{Literature Review}
\label{litrev}

%In this section, we study the use of deep learning methods for the segmentation of open-space.

%--------------------------- Literature review of Segmentation

Recent work using deep learning for semantic segmentation can use two approaches to generate a refined mask: using an \textit{encoder-decoder} architecture to recover fine segmentation predictions~\cite{FCN}, and through the use of \textit{atrous convolutions} to avoid decimating the input's resolution~\cite{2016_deeplabv2}. The former's compuational cost can vary depending on the decoder's architecture, while the latter comes with a sizeable increase in memory footprint, due to the use of large feature maps in the network. \textit{Encoder-decoder} architectures typically make use of an image classification network as its \textit{encoder}, while building a \textit{decoder} to recover fine features for segmentation. \textit{Fully Convolutional Network} (FCN)~\cite{FCN} uses transposed convolutions to upsample the output of the encoder and concatenation with low-level features from the encoder to generate a refined segmentation mask.
%SegNet~\cite{segnet} built on this idea by using unpooling layers for upsampling, as opposed to transposed convolutions.
The use of atrous convolutions for segmentation was pioneered in the DeepLabv2 network~\cite{2016_deeplabv2} by employing varying rates for extracting features and segmenting object at different scales.
%\cite{pspnet} expanded on this idea by using pyramid pooling at the end of the network to gather both global and local information. 
\cite{2018_deeplabv3+} adds a small decoder to a \textit{DeepLabv2} encoder.

%--------------------------- Literature review of DATASETS

Currently, there are a few public datasets available for ADAS applications that, to some extent, include radar information~\cite{nuscenes2019, oxfordRadar2019}. The radar \textit{Robotcar} dataset~\cite{oxfordRadar2019} was released for scene understanding analysis with radar data. The dataset includes radar, lidar, camera, GPS, and IMU observations. The radar data was collected using a special purpose sensor with much higher resolution and range than average industrial radar and is not designed based on the requirements of the automotive industry. %Collected data is a $360$ degree azimuth-range representation of the received power reflection. 
Although this dataset does not provide the raw signal, azimuth-range representation still provides more insight into the radar data than the detected points of \cite{nuscenes2019}. The major disadvantage of this dataset is the lack of ground-truth bounding box information for the objects at the time of writing of this paper.
%Another publically available dataset which contains RADAR information is the FieldSafe dataset. Although this dataset is for agricultural applications, it still conveys some information that could be useful to analyze. The dataset comprises of approximately 2 hours of raw sensor data from tractor-mounted sensor system in a grass mowing scenario. Sensor modalities include a stereo camera, thermal camera, web-camera, 360-degree camera, lidar, and radar. Localization ground-truth is extracted by fusion of IMU and GNSS. The annotations for static and moving obstacles, including humans, mannequin dolls, rocks, barrels, buildings, vehicles, and vegetation, are presented in global coordinate frames.
Annotating the radar data is especially challenging. It is hard to understand the information as displayed in common representations. 
%Using detected point clouds, which apply a detection method on the data, causes the loss of some of the radar data's information. 
To address this issue, one can rely on the use of a visual or depth estimation sensor in combination with radar to create the ground truth data. %As radar is a depth estimation sensor, calibrating it with other depth estimation sensors is a trivial task. However, extrinsic calibration of radar with camera is a much more difficult process. 

%--------------------------- Literature review of Deep Radar

Most radar processing currently uses traditional techniques. In the case of occupancy grid mapping, it is common to use \textit{Inverse Sensor Models} (ISM), followed by Bayesian filtering. Sless et al.~\cite{sless2019road} proposes a U-Net inspired segmentation architecture which takes a Bird's Eye View (BEV) input and generates a mask containing a prediction for each pixel: occupied, unoccupied or unobservable. Formulating the problem as a three-class segmentation problem shows an important improvement when compared to traditional methods. This is expanded by \cite{2019_deepISM} by adding of a fourth, \textit{unknown}, class. This latter approach relies more heavily on the certainty of the predictions.

%--------------------------- ENDS

%%%%%%%%%%%%%%%%%%%%%%%%%%%%%%%%%%%%%%%%%%%%%%%%%%%%%%%%%%%%%%%%%%%%%%
%%%%%%%%%%%%%%%%%%%%%%%%%%%%%%%%%%%%%%%%%%%%%%%%%%%%%%%%%%%%%%%%%%%%%%
%%%%%%%%%%%%%%%%%%%%%%%%%%%%%%%%%%%%%%%%%%%%%%%%%%%%%%%%%%%%%%%%%%%%%%
%%%%%%%%%%%%%%%%%%%%%%%%%%%%%%%%%%%%%%%%%%%%%%%%%%%%%%%%%%%%%%%%%%%%%%

\section{Dataset}
\label{dataset}
To the best of our knowledge, there are no publicly available datasets for radar with accessible Analog-to-Digital Converter (ADC) signals and annotations. To overcome this problem, we collect our own dataset. We moved a car equipped with a side view radar and camera in a parking lot with the objective of identifying the drivable open spaces in the scene. A Linear Frequency Modulation Continuous Wave (LFMCW) radar with 76~Ghz frequency in Multiple Input Multiple Output (MIMO) mode is utilized to collect the environment observations. The usage of the Time Division Multiplexing (TDM) MIMO mode results in 8 virtual channels for the radar information: 2 Tx elements transmit sequentially and 4 Rx elements receive coherently. The resulting dataset is made up of  3913 frames, collected in 11 driving sequences.

To collect any radar data, the first step is to select a set of parameters for the signal. Table \ref{tab:seg_params} shows the details of the configuration used to capture the data.

\begin{table}[h]
	\centering
	\caption{Details of the configuration used with radar.}
	\label{tab:seg_params}
	\begin{tabular}{|l|c|}
		\hline
		Frequency            &     76 \textit{Ghz}           \\\hline
		Maximum Range   	 & 	   15 \textit{m}             \\\hline
		Range resolution  	 &     0.12 \textit{m}           \\\hline
		Unambiguous Velocity &     10.5 \textit{m/s}         \\\hline
		Velocity resolution  &     0.33 \textit{m/s}       	\\\hline
		Field of View    &     90\degree          \\\hline
	\end{tabular}
	%\medskip
	%\caption{Details of the configuration used with radar.}
	%\textbf{}
\end{table}

A camera is also used in our data capture to assist with the annotation of the data. We fix it to the same frame-rate as the radar and capture images of size 1280x960 pixels. The captured visual information by camera will later be used to create the ground truth labels. All communication between various components and their synchronization is managed through the \textit{Robot Operating System (ROS)} software.%\footnote{www.ros.org}.

\subsection{Radar Processing Pipeline}
%Multiple chirps are received by the Rx elements of the radar. 
The signals at each Rx element have the same amplitude but different phase values that represent angular spectrum once converted to frequency domain. The radar echoes are transformed to a 3D tensor of Samples-Chirps-Antennas (SCA). The SCA tensor consists of complex numbers. This is the earliest level to process the radar information. At this stage, all the information is in the time domain and there is no spatial coherence between the values. This entails that any model applied to this stage should explicitly or implicitly include layers to extract spatial coherency.

\begin{figure}[h]
	\begin{center}
		\includegraphics[width=0.9\linewidth]{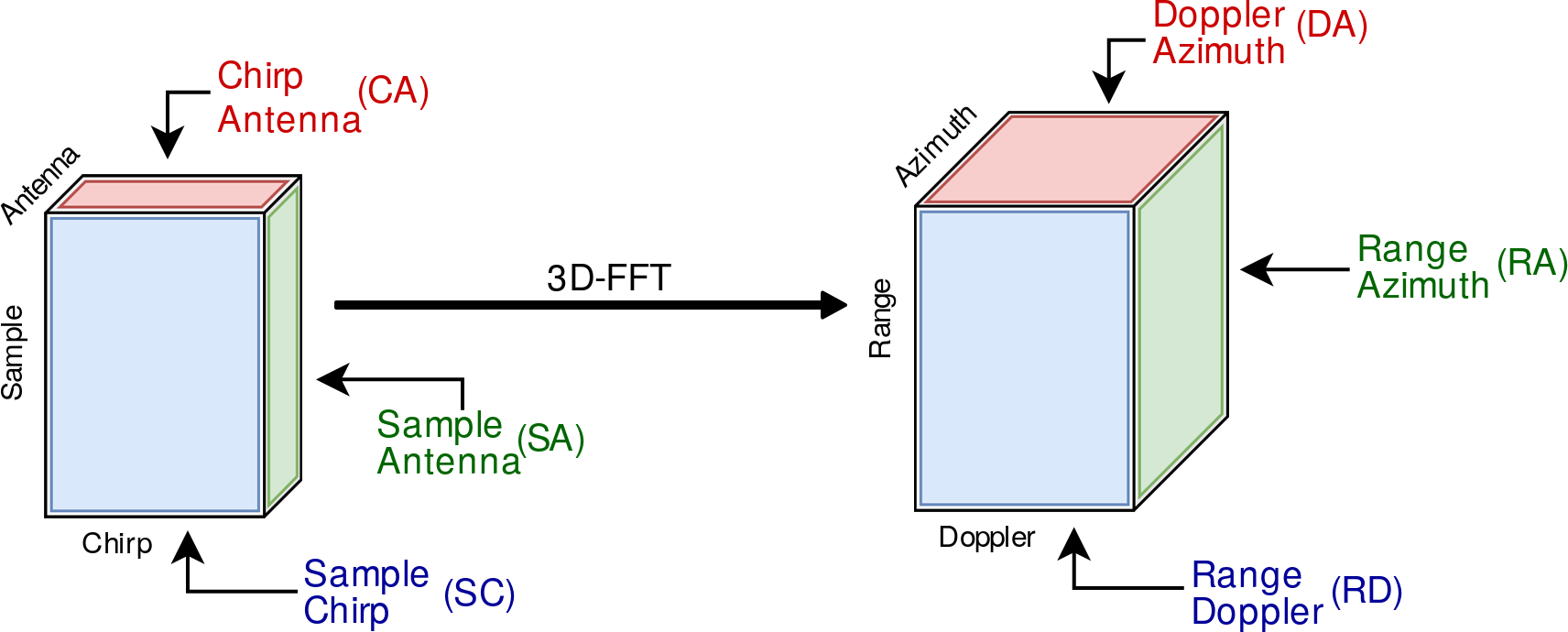}
	\end{center}
	\caption{Radar data representation. (Left) SCA representation. Arranging data along antennas results in Sample-Chirp (SC) view, arranging along chirps results in Sample-Antenna (SA), and Chirp-Antenna (CA) is achieved by arranging data along sample dimension. (Right) Fourier transformation is applied along all three dimensions. Arranging values along azimuth results in Range-Doppler (RD), arrangement along doppler results in Range-Azimuth (RA), and arrangement along range results in Doppler-Azimuth (DA).}
	\label{fig:OD_RA_DA_RD}
\end{figure}

Applying Fast Fourier Transform (FFT) along Samples, Chirps, and Antenna dimension results in Range-Doppler-Azimuth (RDA) representation. %Please note that prior to applying the FFT along the Antenna dimension, the tensor is appended with zeros to a desired tensor resolution.

Range-Azimuth (RA) is the spatial representation of the received signals. It represents the Bird-Eye-View (BEV) of the environment in polar coordinates. A Polar to Cartesian transformation is regularly used on this representation to calculate the direction of arrival (DoA) point-cloud map in Cartesian coordinates for detected objects. In our dataset, we store SCA, RDA and DoA tensors to cover all the various mainstream levels of inputs to any system. 

%We propose three different modes to consume radar data in our deep model. This includes using 3D RDA tensor, utilizing RA representation, and using the Cartesian DoA map.

\subsection{Annotation Challenge}
\label{annos_subsec}
Annotating radar data is an extremely challenging task as the echo-responses are not easily understandable for human. DoA point cloud is an easier representation to understand, but the level of information is coarse, such that it is extremely difficult to use for annotation. To overcome this issue, we employ the sequence of monocular camera images collected in synchronization with radar. As the calibration of a single monocular camera to the radar sensor is prohibitively difficult, we rely on scene reconstruction techniques to extract odometry information. Once an odometry trace is calculated, we use this trace to accumulate the radar DoA's. The open source software of \cite{Meshroom-1} is used to perform 3D reconstruction and extract the odometry trace.

%The accumulated DoA provides rich enough information to annotate with ease, and requires minimal frame based alterations. Later, these annotations are propagated to corresponding frames in an automatic manner, based on the odometry and radar intrinsic parameters. This ground truth generation pipeline is shown in Figure \ref{fig:gt_pipeline}.

\begin{figure}[h]
	\begin{center}
		\includegraphics[width=0.99\linewidth]{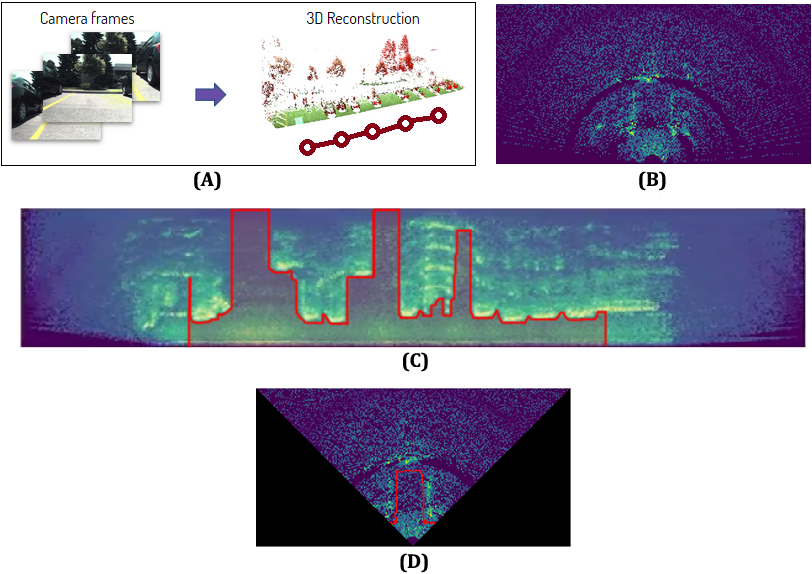}
	\end{center}
	\caption{Ground truth generation pipeline. (A) Camera frames are used to reconstruct the scene and extract odometry. (B) Cartesian DoA from a single frame. (C) Accumulated Cartesian DoA which is used to generate initial annotations. (D) Annotations are distributed to the corresponding frames, cropped for the radar's field of view, and are manually adjusted.}
	\label{fig:gt_pipeline}
\end{figure}

Open space annotations from accumulated DoA are propagated to corresponding frames using the odometry and radar intrinsic parameters. This ground truth generation pipeline is shown in Figure \ref{fig:gt_pipeline}. 

As the annotations are propagated from 3D reconstruction, labels include values for locations that are not in the direct line of sight of the radar. Even though a radar can still detect free space in those regions, they are removed to be consistent with the single frame based annotations of \cite{oxfordRadar2019}.

Note that the inputs to our proposed models only include the radar data, and the output is generated as the occupancy grid map in Cartesian coordinates. Also, it is worth mentioning that the annotation task can be handled much more easily if a laser depth sensor such as Lidar would be used.

The SCORP dataset is available at this link~\footnote{www.sensorcortek.ai/publications/}.

\section{Model Architecture}
\label{model}
We propose and compare three deep learning approaches inspired by recent work in the field of semantic segmentation: \textit{DeepLabv3+}~\cite{2018_deeplabv3+},  \textit{Fully Convolutional Networks}~(FCN)~\cite{FCN} and \textit{FCN\_tiny}. We implement all three of these segmentation models with MobileNet-v2 as their feature extractor.

%MODEL PROPOSAL GOES HERE

%%%%%%%%%%%%%%%%%%%%%%%%%%%%%%%%%%%%%%%%%%%%%%%%%%%%%%%%%%%%%%%%%%%%%%
%To compare our proposed model, we have implemented two of the state-of-the-art, DeepLabv2 \cite{2016_deeplabv2} and FCN \cite{FCN}.

The DeepLabv2 architecture for segmentation is implemented through two core concepts: reducing the output stride of the feature extractor while using atrous convolutions to generate larger feature maps, and performing atrous spatial pyramid pooling (ASPP) to cover a wider range of object sizes. The \textit{DeepLabv3+} architecture iterates on the architecture of the ASPP, and appendes a small decoder to the DeepLabv2 encoder network, which upsamples the feature extractor's output and combines them with features from earlier layers. We implement a complete version of the DeepLab architecture, referred to as \textit{DeepLabv3+} in our experiments. This version uses atrous convolutions of rates 2, 4 and 6 in its ASPP module. The model's output is then resized to the input size through bilinear interpolation.

%\begin{figure}[t]
%	\begin{center}
%		\includegraphics[width=\linewidth]{model_deeplabv3.png}
%	\end{center}
%	\caption{DeepLabv3+ model architecture with MobileNetv2 of output stride 16 as a feature extractor.}
%	\label{fig:model_deeplabv3}
%\end{figure}

%%%%%%%%%%%%%%%%%%%%%%%%%%%%%%%%%%%%%%%%%%%%%%%%%%%%%%%%%%%%%%%%%%%%%%
The FCN segmentation architecture is a simple encoder-decoder method which uses transposed convolutions to upsample feature maps by a factor of 2. The encoder's output is upsampled and concatenated with lower level feature maps, thereby recovering detailed spatial information. Two of these upsampling steps are used, generating output feature maps 8 times smaller than the network's input. These are then resized to the input size through bilinear interpolation.

Finally, we experiment with a small variation of the FCN architecture, FCN\_tiny, with a reduced number of feature maps in each MobileNetv2 layer by 75\%, and using a depth of 8 in the decoder feature maps. The resulting model has a much lower number of parameters than the other models.

%\begin{figure}[t]
%	\begin{center}
%		\includegraphics[width=\linewidth]{model_fcn.png}
%	\end{center}
%	\caption{FCN model architecture with MobileNetv2 of output stride 32 as a feature extractor.}
%	\label{fig:model_fcn}
%\end{figure}

%%%%%%%%%%%%%%%%%%%%%%%%%%%%%%%%%%%%%%%%%%%%%%%%%%%%%%%%%%%%%%%%%%%%%%
%%%%%%%%%%%%%%%%%%%%%%%%%%%%%%%%%%%%%%%%%%%%%%%%%%%%%%%%%%%%%%%%%%%%%%
%%%%%%%%%%%%%%%%%%%%%%%%%%%%%%%%%%%%%%%%%%%%%%%%%%%%%%%%%%%%%%%%%%%%%%

\section{Experiments}
\label{exps}
We perform a series of experiments that address various aspects of application of deep learning to the radar sensor. We used three distinct data representation, namely, RAD, RA, and DoA. The combination of these representations as input, Polar and Cartesian outputs, along with the model architectures outlined in previous section, results in our list of experiments. We further evaluate the effect of implicit vs explicit Polar to Cartesian coordinate transformation. We use 3193 frames for training, and 720 for evaluation (9 and 2 driving sequences, respectively).

Mean Intersection-over-Union (Mean-IoU), commonly used for semantic segmentation tasks~\cite{FCN}~\cite{2016_deeplabv2}, is selected as the evaluation metric. Mean-IoU is calculated as the average of the Intersection-over-Union (IoU) metric of each class.

%%%%%%%%%%%%%%%%%%%%%%%%%%%%%%%%%%%%%%%%%%%%%%%%%%%%%%%%%%%%%%%%%%%%%%
\subsection{Input Modalities}
The goal of this experimentation is to identify the effect of various input and output representations on the model performance. We compare three distinct input modalities:

%%%%%%%%%%%%%%%%%%%%%%%%%%%%%%%%%%%%%%%%%%%%%%%%%%%%%%%%%%%%%%%%%%%%%%
\begin{itemize} 
	\item \textbf{RAD:} RAD is a 3D tensor and the convolutions are applied along the last (Doppler) dimension. This input is the polar representation of radar frames for each individual Doppler channel that is generated from the third FFT along antenna dimension of SCA. 
	
	\item \textbf{RA:} RAD input Tensor is summed along the Doppler dimension and the logarithmic value of the matrix is named as RA. This representation is the actual value that traditional methods use to extract the location of their detections. Same as RAD, information in RA is also in polar coordinates.
	
	\item \textbf{DoA:} DoA input matrix is a Cartesian Bird-Eye-View (BEV) generated from RA. The pixel values in the matrix represent the power received by the radar sensor at that location. The benefit of this modality is its metric coherence with convolutional kernels. This is important as the same convolutional filter in every location of this representation represents the same receptive field.
\end{itemize}

In order to use segmentation results in various tasks, they need to be in the Cartesian coordinate system. However, the predictions in polar coordinates can be simply converted to Cartesian system. To isolate the effect resulting from having annotations in two different domains, and compare the Polar inputs to Cartesian inputs fairly, we utilize two output representations:

\begin{itemize} 	
	\item \textbf{Cartesian ground-truth:} As discussed in section \ref{annos_subsec}, the open-space is annotated in a parking lot using DoA-input Tensor. Then, the annotated points are used to generate a mask for open-space segmentation. However, we confined the field-of-view of radar to 90\degree. Cartesian ground truth can be used with all three input models.
	
	\item \textbf{Polar ground-truth:} After generating the Cartesian ground-truth masks, the annotated points are transformed into the Polar coordinate system. For training, we cropped the RA and RAD input tensors to match the selected field of view. %Note that cropping along columns (angle dimension) equates to reducing the angular field of view of the tensor in the Cartesian domain.
\end{itemize}

We conducted experiments where the input tensor is in one domain (i.e. Polar), while the output mask is in another domain (i.e. Cartesian). We expect that the model architecture should learn to adapt the transformation and generate comparable results. Table \ref{tab:experimental-results} shows the results of these experiments. As expected, having the input and output in the same domain in all cases resulted in better performance than learning the domain transformation internally.

We can further observe that using RAD as the input provides the best mean-IOU. This outcome is due to the descriptive information present in RAD that are manually summarized in RA and DoA representations. It is apparent that the model is extracting a better mid-level representation than the manual compression achieved through summation or coordinate transformation done by RA and DoA. RA beats the DoA in performance. This shows that a model using convolutional kernels defined with Cartesian coordinate in mind, is capable of adapting them to the Polar usecase. From a sensor point of view, far points in the BEV map of DoA have much lower information density compared to the closer points. This imbalance is a reason for the lower performance of the DoA input.

\begin{table}[h]
	\centering
	\caption{Mean-IoU reported for different model architectures.}
	\label{tab:experimental-results}
	\resizebox{0.475\textwidth}{!}{%
		\begin{tabular}{|l|l|l|l|l|}
			\hline
			Input  & Label                                &      FCN\_tiny                                      &                   FCN                &               DeepLabV3+                \\ \hline
			RAD-Input       & RA-Mask                                       & \multicolumn{1}{l|}{83.61}                 & \multicolumn{1}{l|}{83.76}        & \multicolumn{1}{l|}{82.88}               \\ \hline
			RAD-Input       & DoA-Mask                                      & \multicolumn{1}{l|}{73.24}                 & \multicolumn{1}{l|}{78.05}        & \multicolumn{1}{l|}{73.92}               \\ \hline
			RA-Input        & RA-Mask                                       & \multicolumn{1}{l|}{81.99}                 & \multicolumn{1}{l|}{82.59}        & \multicolumn{1}{l|}{81.14}               \\ \hline
			RA-Input        & DoA-Mask                                      & \multicolumn{1}{l|}{77.96}                 & \multicolumn{1}{l|}{78.24}        & \multicolumn{1}{l|}{77.22}               \\ \hline
			DoA-Input       & DoA-Mask                                      & \multicolumn{1}{l|}{79.00}                 & \multicolumn{1}{l|}{80.75}        & \multicolumn{1}{l|}{78.05}               \\ \hline
		\end{tabular}%
	}
	%\medskip
	%\caption{Mean-IoU reported for different model architectures based on our experiments.}
	%\label{tab:experimental-results}
\end{table}

%%%%%%%%%%%%%%%%%%%%%%%%%%%%%%%%%%%%%%%%%%%%%%%%%%%%%%%%%%%%%%%%%%%%%%
\subsection{Model Analysis}
\label{modelanalysis}
%Our experiments are based on training the three model architectures mentioned in section \ref{model_arch} with different input modalities. The main goal of these experiments is to validate the generalization of the model architecture in two domains (Cartesian and Polar). Our entire list of designed experiments along with their evaluation results on three model architectures are mentioned in table \ref{tab:experimental-results}.

%\footnote{www.tensorflow.org}
\textit{Tensorflow} is used as a back-end platform for training and evaluation of these models. We used \textit{Trainable Softmax Cross-Entropy loss} with initial learning rate of 0.005. This loss is based on Softmax Cross-Entropy loss, with the addition of trainable parameters used to weight each class during loss calculation, thereby avoiding the need to hand-pick appropriate weighting parameters. The optimizer used is RMSProp with a momentum and a decay factor of 0.9. All training of our model architectures was undertaken on \textit{Nvidia Geforce RTX 2080 TI}. Based on our experiments, we noticed that all of our implemented model architecture reached to their optimized weights within 22-30K steps.

Training results show that the FCN model is clearly the better model. In all cases, it achieves superior performance than its opponents. FCN\_tiny, a compact version of FCN with only $210k$ parameter - less than 10\% of FCN's size - performs comparably to the much larger DeepLabV3+ model. This shows that atrous convolutions are not as helpful as skip architectures for this task. This is most likely caused by the local nature of useful information for this task, which reduces the need for larger atrous rates. Downsampling the input by a factor of 32 provides sufficient information without significantly affecting the quality of the features. On the other hand the holes in an atrous convolution hampers the use of local information.

The proposed FCN\_tiny model achieves $324$ FPS on the \textit{RTX 2080 TI} GPU and $267$ FPS on the \textit{Intel Core i9-9900K} CPU. As such, it is $8\%$ faster than FCN on GPU ($301$ FPS) and more than twice as fast as FCN on CPU ($118.56$ FPS).

\section{Conclusion}
\label{conclusion}
In this paper, we evaluated various representations of radar data as inputs to deep models, various deep model architectures, and the effect of the polar to Cartesian transformation. 
%It is shown that the FCN model performs better than the DeepLabV3+ model for this task.
FCN\_tiny has slightly worse performance than FCN while being an order of magnitude smaller, making it the perfect candidate for Radar-on-Chip integration. To the best of our knowledge, SCORP constitutes the first comprehensive dataset that provides ADC information.

%Our experiments only covered single frame observations that do not employ any temporal information, keeping the computational requirements as low as possible. 
Employing temporal models would increase the performance of occupancy map predictions. We keep this aspect as a topic to address in our future research.
We hope this paper and the new dataset will provide an insight into the inner workings of the radar sensor, and enable an increasing number of researchers to easily access the radar data and further develop this field.

%\vspace{12pt}
%\color{red}
%IEEE conference templates contain guidance text for composing and formatting conference papers. Please ensure that all template text is removed from your conference paper prior to submission to the conference. Failure to remove the template text from your paper may result in your paper not being published.

\end{document}